# Absence of anomalous underscreening in highly concentrated aqueous electrolytes confined between smooth silica surfaces

Saravana Kumar[1], Peter Cats[2], Mohammed B. Alotaibi[3], Subhash C. Ayirala[3], Ali A. Yousef[3], René van Roij[2], Igor Siretanu[1], Frieder Mugele[1].

1. Physics of Complex Fluids Group and MESA+ Institute, Faculty of Science and Technology, University of Twente, PO Box 217, 7500 AE Enschede, The Netherlands.
2. Institute for Theoretical Physics, Center for Extreme Matter and Emergent Phenomena, Utrecht University, Princetonplein 5, 3584 CC Utrecht, The Netherlands
3. The Exploration and Petroleum Engineering Center – Advanced Research Center (EXPEC ARC), Saudi Aramco, Dhahran 34465, Saudi Arabia

**Abstract**

Recent experiments and a series of subsequent theoretical studies suggest the occurrence of universal underscreening in highly concentrated electrolyte solutions. We performed a set of systematic Atomic Force Spectroscopy measurements for aqueous salt solutions in a concentration range from 1 mM to 5 M using chloride salts of various alkali metals as well as mixed concentrated salt solutions (involving both mono- and divalent cations and anions), that mimic concentrated brines typically encountered in geological formations. Experiments were carried out using flat substrates and submicrometer-sized colloidal probes made of smooth oxidized silicon immersed in salt solutions at pH values of 6 and 9 and temperatures of 25 °C and 45 °C. While strong repulsive forces were observed for the smallest tip-sample separations, none of the conditions explored displayed any indication of anomalous long range electrostatic forces as reported for macroscopic mica surfaces. Instead, forces are universally dominated by attractive van der Waals interactions at tip-sample separations of $\approx 2$ nm and beyond for salt concentrations of 1 M and higher. Complementary calculations based on classical density functional theory for the primitive model support these experimental observations and display a consistent decrease in screening length with increasing ion concentration.

**Introduction**

The development of Derjaguin-Landau-Verwey-Overbeek (DLVO) theory of electrostatic and van der Waals interactions in electrolytes is one of the highlights of colloid science in the 20[th] century [1-4]. It has been very successful at describing colloidal stability, charged interfaces, and interaction forces at sufficiently large distances and low ion concentrations. However, being a theory based on point ions in a solvent continuum, it is not applicable at separations and concentrations where the finite size of ions and solvent molecules can no longer be neglected [5, 6]. Under these conditions, various microscopic properties of the system need to be taken into account, including the size of ions and molecules, ion-specific interaction forces, polarizability, correlation effects, and solvation. While many of these non-DLVO interactions have been analyzed in detail in theory, disentangling the different contributions in experiments, typically at distances of 1-2nm or below, has proven extremely difficult and has therefore become a field of intense investigation and debate. Despite this difficulty, our knowledge about solid-liquid interfaces has increased dramatically thanks to developments in various experimental and numerical techniques, including scanning force microscopy, (surface) x-ray diffraction, non-linear optical spectroscopy, molecular dynamics simulations, as well as density functional theory [7-12]. Numerous studies have revealed detailed insights, amongst others, into the adsorption of ions and the hydration of various solid surfaces, in particular including atomically flat crystalline mica and (almost) similarly smooth but amorphous silica surfaces.

In recent years, highly concentrated electrolytes have re-gained additional attention because of their enormous relevance in many fields of science and engineering such as biology, geology, oil recovery, storage of renewable energy in batteries and capacitive devices, etc. [13-18]. A particularly spectacular observation has been the reports [19-23] of monotonically decaying forces of apparently electrostatic origin with a range of several nanometers for ion concentrations of the order of 1 M and higher. In contrast to conventional wisdom based on Debye screening, the decay length of these forces at these high concentrations was found to increase with increasing ion concentration. These observations were initially based on force measurements on confined ionic liquids with a strongly layered arrangement close to the solid surface [24, 25] and they were, more recently, generalized to highly concentrated aqueous salt solutions [20, 21]. To rationalize the observations, it was proposed [21] to consider the highly concentrated electrolyte as an ensemble of densely packed neutral ion pairs that obtains a finite conductivity by the introduction of defects consisting of unpaired ions. In this scenario, the reduction of ion concentration increases the chance of forming such defects and therefore improves screening leading to a shorter screening length. Vice versa, increasing the salt concentration would lead to a reduction of free charges and thereby increase the decay length of electrostatic interactions. Specifically, this proposition leads to a supposedly universal scaling law for the screening length $\lambda_s \propto l_B d^3 \rho$, where $l_B$ is the Bjerrum length, $d$ the average ion diameter, and $\rho$ the salt concentration. This effect has been denoted as anomalous underscreening.

The experimental observations in Refs.[19, 21, 22, 25] sparked a renewed interest in older theoretical models of underscreening in concentrated electrolytes [26, 27]. On the one hand the more recent calculations indeed confirm that the decay length of the force does increase with increasing ion concentration beyond the so-called Kirkwood point, where the interaction force changes from monotonically decaying with distance to oscillatory (with an exponentially decaying amplitude). On the other hand, however, the absolute increase of the decay length in the calculations has in all cases considered been much more moderate than the one reported in Refs. [20, 21]. Moreover, beyond the Kirkwood point, by definition, the theory predicts a decaying oscillatory force profile, whereas the decay lengths reported in the underscreening experiments exhibit a monotonic exponential force profile. It has also been argued that the ion-depairing model of Ref.[21] is much less plausible for concentrated aqueous salt solutions that typically still display a substantial excess of solvent molecules as compared to pure ionic liquids [28]. While the theoretical models refer to intrinsic properties of the concentrated electrolytes in the bulk, the experiments are based on force measurements between two solid surfaces that are separated by thin films of electrolyte with a thickness of a few nanometers. Given the large surface-to-volume ratio of these systems, it is not obvious whether the forces measured under these conditions are indeed dominated by intrinsic properties of the fluid or to what extent they are affected by properties of the specific solid-liquid interfaces. In general, however, the strength (i.e., the amplitude) of the force does depend on the properties of the solid-liquid interfaces while their asymptotic range (decay length) is only a bulk property of the electrolyte.

In this work, we conduct an independent study where we measure the interaction forces at silica-brine interfaces using colloidal probe atomic force microscopy (AFM). The bulk salt concentration is varied over more than three orders of magnitude and the forces are measured at different temperatures and pH. These experimental conditions help us put the proposed scaling law [21] to test by changing the Bjerrum length (changed via both temperature and bulk dielectric constant). We also vary the ion size, which is another important ingredient of the scaling law. In addition to measuring interaction forces in simple alkali halide solutions, we also measured interaction forces in complex brines such as high salinity water (HSW), which mimics the composition of sea water, and rock-formation water (FW). Our measurements are complemented by classical density functional theory (DFT) calculations using the primitive model (PM) with both hydrated and bare ion diameters in a confined geometry, which allow for a direct comparison between experimental and theoretical decay lengths. Neither experiments nor calculations display any dramatic increase in decay length at salt concentrations beyond 1 M.

**Materials and Methods**

*Experimental*

Experiments were carried out using 1x1 cm$^2$-sized pieces of silicon wafers (resistivity of $1-5$ Ωcm) with a 30 nm thick thermally grown oxide layer. Prior to every experiment, the samples were cleaned in a 50:50 mixture of absolute ethanol and absolute isopropanol (ACS, Sigma Aldrich) for 20 min., blown dry in a stream of N$_2$ gas and subsequently plasma cleaned for 20 min. The samples were then glued to a magnetic sample puck using UV-curable Norland optical adhesive 81 (Norland Products, Inc.).

AFM experiments were performed in an Asylum Research Cypher AFM equipped with a closed environmental cell (ES) for measurement in liquid under controlled gas atmosphere. Hemispherical colloidal AFM probes (LRCH, Team Nanotec) made of silicon with a thin native oxide layer having a nominal tip radius of $R = 250$ nm and nominal cantilever spring constant of $k_c = 0.7$ N/m were used for all experiments. The actual values of $R$ and $k_c$ were obtained independently for each probe using scanning electron microscopy (see Figure S1) and the thermal tune method [29], respectively. The tip-apex (100 x 100 nm$^2$) of the colloidal probe has a root-mean-square (rms) roughness value of 0.176 nm (see Figure S2). The AFM probes were subjected to the same cleaning procedure as the samples.

All salt solutions were prepared freshly prior to each experiment by dissolving reagent-grade salts (ACS, Sigma Aldrich) in demineralized water (milli-Q; conductivity 18.2 mΩcm). Measured pH values and electrical conductivities of all electrolytes are given in the Table S1 The composition of the more complex brines HSW and FW, which contain multiple mono- and divalent cations and anions are specified in Table S2. Sample and probe were mounted in the AFM liquid cell in the presence of ∼ 200 μL of salt solution. Fluids within the cell were exchanged by manually injecting at least 3 mL of the new solution into the cell while simultaneously sucking out the old one to keep the total fluid volume constant. For experiments at elevated pH, NaOH was added to the salt solution immediately before injecting into the AFM cell and the residual gas volume in the cell was filled with $N_2$ gas to minimize $CO_2$-induced pH reduction. Experiments were carried out at temperatures of $T = 25 \pm 0.1°C$ and $45 \pm 0.1°C$, as set by the integrated temperature controller of the AFM. For experiments at 45 °C, the area surrounding the sample stage inside the AFM cell is filled with demineralized water to minimize evaporation of the experimental solutions from the sample stage.

The AFM was operated in static mode. At least 100 deflection vs piezo displacement curves were recorded for each condition at a displacement rate of 60 nm/s at an acquisition rate of 0.5 Hz. A maximum deflection threshold of ≈ 6 nm was chosen to guarantee hard tip-sample contact at maximum deflection (i.e., a one-to-one correspondence between piezo displacement and cantilever deflection). Unless noted otherwise, only data recorded upon approach will be presented. The raw data were transformed to force vs tip-sample separation curves using standard procedures (see Figure S3). To eliminate the thermally induced vibration of the cantilever (see grey symbols in FigureS3) and to speed up numerical data analysis, the density of data points was reduced by using a moving box-average of 0.1 nm width with a 50% overlap between adjacent points. Prior to averaging, force curves obtained under identical conditions were aligned in the (normal) z-direction using the point of maximum cantilever deflection, i.e., the deflection threshold mentioned above.

To obtain the surface charge densities, the experimental force-separation curves were fitted with theoretical DLVO force curves that has contributions from the electrostatic interaction $F_{el}$ and van der Waals interaction $F_{vdW}$,

$$F_{DLVO} = F_{el} + F_{vdW} \quad (1)$$

The electrostatic part was obtained by solving the full Poisson-Boltzmann equation with a boundary condition that involves a constant regulation [30-32]. A detailed description of the surface charge extraction from the electrostatic interactions is given in the supplementary material . The van der Waals' contribution is calculated as the interaction between a plane-sphere geometry using the formula [3],

$$F_{vdw} = -\frac{A\,R}{6\,(H+\delta)^2} \qquad (2)$$

where $H$ is the tip-substrate separation, $A$ is non-retarded Hamaker constant, and $\delta \ll 1$ nm is a small empirical offset parameter to regularize the divergence of the force for $H \to 0$. According to literature, $A$ assumes values of $1.6 \ldots 6 \times 10^{-21} J$ [33-36] for a SiO$_2$-aqueous NaCl/water-SiO$_2$ system, where the positive sign implies an attractive van der Waals force.

*Theoretical Calculations*

The DFT calculations in the present work were carried out in the framework of the primitive model, as described in detail in ref.[27]. Briefly, the solvent is treated as a continuous dielectric medium with a bulk concentration-dependent dielectric constant $\varepsilon_r$ (see Table S3) at temperature $T = 298.15$ K. Anions and cations are modelled as centro-symmetric charged hard spheres with ion diameters $d_+$ and $d_-$, respectively. The electrolyte is confined between two planar charged hard walls, separated by a distance $H$, at which a small constant electric potential $\Phi_0 = 25$ mV is applied with respect to a grounded ionic reservoir in osmotic equilibrium, as illustrated in Figure S4. The low applied potential, chosen for convenience, does not affect the asymptotic decay, which is a bulk property of the electrolyte.

In short, DFT involves the grand potential functional, which reads for our system [37]

$$\Omega[\{\rho\}] = \mathcal{F}_{id}[\{\rho\}] + \mathcal{F}_{ex}^{HS}[\{\rho\}] + \mathcal{F}_{ex}^{ES}[\{\rho\}] \\ - \sum_{j=\pm} \int \mathrm{d}\mathbf{r}\, \rho_j(\mathbf{r})\left(\mu_j - V_{ext}^j(\mathbf{r})\right) - Q\Phi_0, \qquad (3)$$

where $\rho_j(\mathbf{r})$ is a variational density profile, $\mu_j$ the chemical potential, and $V_{ext}^j(\mathbf{r})$ the external potential that acts on the two ionic species labeled by $j = \pm$. The total charge $Q$ is that on the two charged surfaces and $\Phi_0$ is the electric potential that is applied to the electrodes with respect to a grounded ion reservoir. The first term $\mathcal{F}_{id}[\{\rho\}]$ is the Helmholtz free-energy functional for an ideal (non-interacting) system, the second term $\mathcal{F}_{ex}^{HS}[\{\rho\}]$ accounts for the steric hard-sphere interactions and is dealt with by Fundamental Measure Theory White-Bear II [38, 39], and the third term $\mathcal{F}_{ex}^{ES}[\{\rho\}]$ describes the Coulombic interactions for which we invoke a functional using the direct correlation function obtained from the analytic solution of the Ornstein-Zernike equation within the well-known Mean Spherical Approximation [27, 40, 41].

At fixed chemical and external potentials, the grand potential functional $\Omega[\rho]$ is minimized by the equilibrium density profiles $\rho_0$, i.e.

$$\left.\frac{\delta\Omega[\{\rho\}]}{\delta\rho_j(\mathbf{r})}\right|_{\{\rho\}=\rho_0} = 0. \qquad (4)$$

Moreover, the grand potential at its minimum is the equilibrium grand potential of the system at the given thermodynamic condition, i.e., $\Omega[\{\rho_0\}] = \Omega(\{\mu\}, V, T, \Phi_0, H)$. Hence, DFT is a powerful theoretical framework to combine thermodynamics and structure as it gives access to the thermodynamics of the system via the equilibrium density profiles, that can be calculated numerically from the Euler-Lagrange (Eq. 4). In particular, DFT gives access to the surface free energy

$$\gamma(H) = \frac{\Omega(H)}{A} + pH \tag{5}$$

where $p$ is the osmotic bulk pressure of the electrolyte and $A$ the fluid-solid interaction area. This surface free energy $\gamma$ is related to the force $F_{DLVO}$ via the Derjaguin approximation [3]

$$\gamma(H) - \gamma(\infty) = \frac{F_{DLVO}(H)}{2\pi R}, \tag{6}$$

with R=250 nm the tip radius introduced before and $\gamma(\infty)$ the surface free energy at large separations when the surfaces do not interact. For relatively large $R$, the convex surface of the tip can be modelled as a planar surface. Our DFT calculations are performed for surface separation ranging from $H = 1$ nm to $H = 80$ nm.

**Results**

Fig. 1 shows an overview of the averaged experimental force curves in NaCl solutions across the full range of concentrations from 1 mM to 5 M. Overall, the curves follow the classical expectations based on DLVO theory. At low concentrations, the force is dominated at large separation by electrostatic repulsion between the two symmetric negatively charged silica surfaces and merges into an even stronger monotonically increasing short-range repulsive force upon approaching contact. In line with earlier force measurements in aqueous systems using colloidal force microscopy, we did not observe any oscillatory forces that would result from a layered arrangement of water molecules or ions for any fluid composition investigated. As expected, based on standard DLVO theory, the electrostatic repulsion becomes progressively screened upon increasing the salt concentration. For concentration of $c = 300$ mM and higher, the force at large separations becomes dominated by attractive van der Waals forces. Upon reducing the tip-sample separation to $\approx 1$ nm or less, the van der Waals attraction gives way again to the same short-range repulsive force as in the case of low salt concentrations.

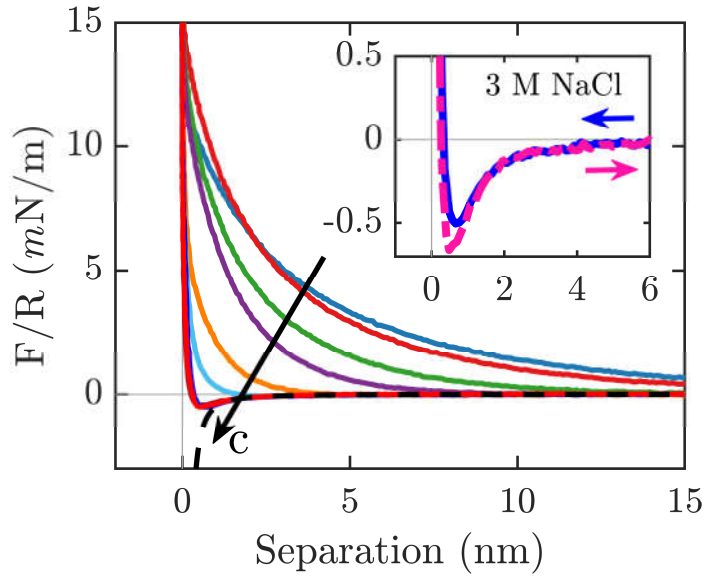

Figure 1: Normalized tip-sample interaction force vs. tip-sample separation for NaCl concentrations of 1 mM, 3 mM, 10 mM, 30 mM, 100 mM, 300 mM, 1 M, 3 M, and 5 M increasing along the arrow as indicated. The temperature $T = 25\ °C$ and the pH is unadjusted. Black dashed line: van der Waals interaction calculated for Hamaker constant $A = 3 \times 10^{-21} J$ and offset (see text) $\delta = 0.1$ nm. Inset: approach and retract force curves for 3 M NaCl.

Qualitatively, our observations are thus in full agreement with the expectations of classical colloid science up to the highest concentrations investigated. We emphasize that this statement equally applies to force curves recorded upon retracting the tip from the sample: for tip-sample separations of $\approx 1.5$ nm and more, approach and retract curves perfectly overlap within the noise level, indicating that the system is thermodynamically equilibrated throughout the measurement. Only for the highest salt concentrations and the smallest tip-sample separations a small degree of hysteresis is observed (see inset Figure 1; see also Figure S3 for an example at low salt concentration).

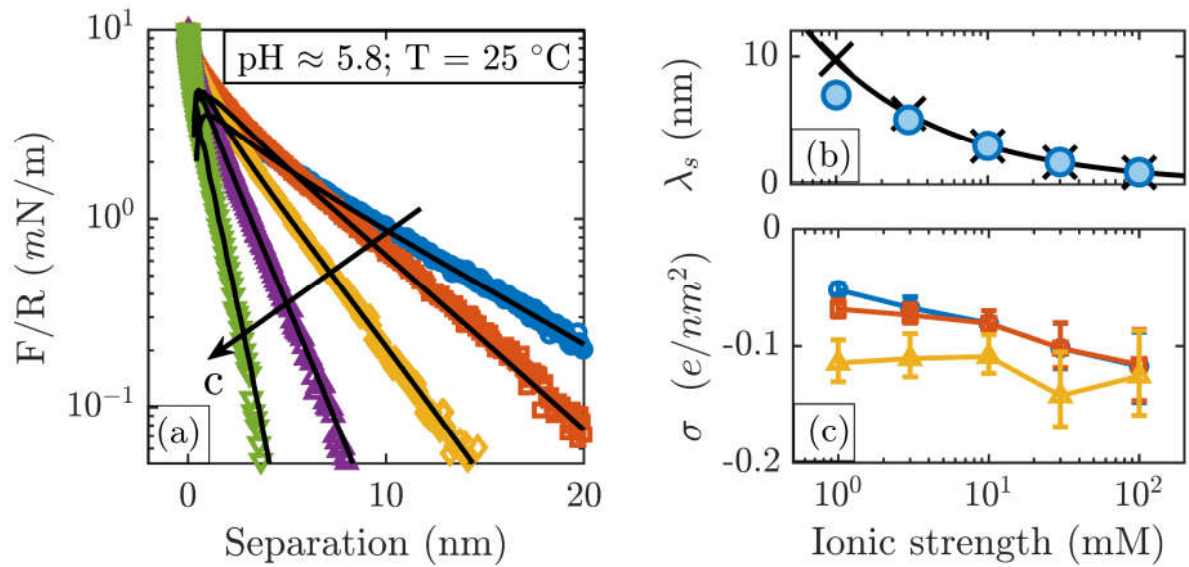

Figure 2: (a) – Normalized interaction forces at silica-NaCl interface for the low-salt concentrations (1 mM, 3 mM, 10 mM, 30 mM, and 100 mM) at 25 °C and a pH ≈ 5.8. The solid black lines indicate the theoretical fit of the DLVO forces, Eq. ( 1 ), to the experimental data (symbols). (b) – The extracted decay length from panel-a (blue symbols) and DFT calculations with bare ion diameters (black crosses) is compared with the theoretical Debye length (solid black line). (c) – Surface charge density of isolated silica surface extracted from the force-distance curves by solving the full Poisson-Boltzmann equation with charge-regulation as boundary condition for various conditions of NaCl solutions: ○ – pH ≈ 5.8 and T = 25 °C; □ – pH ≈ 5.8 and T = 45 °C; ∆ - pH ≈ 9 and T = 25 °C.

**Low-concentration regime** ($c = 1 \ldots 100$ mM)

Figure 2a shows the same normalized total interaction forces as in Figure 1, but on a semi-logarithmic scale for the low concentration regime. The interaction forces show a clear exponential decay at larger separations indicative of electrostatic interactions due to electric double layer overlap. The deviations from purely exponential behavior at small separations are due to charge regulation and other non-DLVO forces such as hydration forces. (For the purpose of this work, we will not make any attempt to identify the detailed physical origin of this short-range force). The decay length of these interaction forces decreases with increasing salt concentration in agreement with the theoretical Debye screening length (black solid line in Figure 2b), and the DFT calculations (black crosses in Figure 2b). The black solid lines in Figure 2a represent fits of DLVO theory to the experimental data using a charge regulation boundary condition in the so-called constant regulation approximation with a regulation parameter *p* between 0 (constant potential) and 1 (constant charge) [30-32]. The resulting surface charge density $\sigma$ at infinite tip-sample separation increases from $\approx -0.05$ to $-0.12 \, e/\text{nm}^2$ upon increasing the NaCl concentration from 1 mM and 100 mM, as shown in Fig. 2c. This increase is in agreement with various earlier measurements at pH 6 [31, 42] and is generally attributed to increased deprotonation of silanol groups enabled by improved screening at higher salt concentration. The values of the regulation parameter p as extracted from the fit (Table S4) indicate that $\sigma$ hardly depends on the tip-sample separation at $c = 100$ mM, corresponding to a constant charge boundary condition ($p \approx 1$). For the lowest salt concentration, a reduction from $-0.5 \, e/\text{nm}^2$ to $-0.3 \, e/\text{nm}^2$ is observed upon bringing the tip to a distance of 2 nm from the surface. This points to a mixed boundary condition ($p \approx 0.6$) with some confinement-induced re-protonation and/or cation adsorption, consistent with earlier reports [31].

Additional force measurements at a slightly elevated temperature of 45 °C and an elevated pH of 9 displayed a qualitatively similar behavior (Figure S5a and b). The resulting surface charge density was found to depend very weakly on temperature (within the narrow range investigated), consistent with earlier streaming potential measurements [43]. At pH = 9.0±0.2 the measurements displayed somewhat more negative surface charges, as expected based on the increased deprotonation of silanol groups at elevated pH [44-47]. In both cases, the decay lengths of the forces were consistent with the ones observed for 25 °C and unadjusted pH within the symbol size in Figure 2b. The DFT predictions lead to somewhat different absolute values of the calculated forces Figure S5c). Yet, more importantly, the observed asymptotic decay lengths agree with the expected Debye screening length (crosses in Figure 2b).

**High-concentration regime** ($c \geq 0.3$ M)

Having verified that the system follows the expectations based on DLVO theory at low concentrations, we turn to the high concentration regime where the Debye screening length is smaller than the average ion diameter, i.e., $\lambda_D < d$. Figure 3 shows the magnitude of the normalized total interaction force for a wide range of ionic strengths, pH, temperature, ion sizes, and brine compositions. For all conditions investigated in the high concentration range, the force is dominated by attractive van der Waals interaction for tip-sample distance of ≈ 2 nm and beyond, as illustrated by the solid black lines with the common slope -2 in double-logarithmic representation (see Figure S6 for a representation of the same data on linear scales). In fact, the solid lines are fits to Eq. 2 with the non-retarded Hamaker constant, $A$, as a fit parameter, which assumes values of $(2.5 - 6.5) \times 10^{-21}$ J, consistent with the expectations of Lifshitz theory. None of the fluids investigated here displays any indication of a significant excess force at distances of 2nm or beyond, irrespective on concentration and ionic species. The latter is particularly apparent in Figure 3f, where we compare chloride salts of Li+, Na+ and Cs+ with widely varying ion diameters of 0.188 nm, 0.234 nm, and 0.372 nm, respectively, which should lead to a strong variation $\propto d^3$ of the observed screening length according to the proposed underscreening model [21]. (NOTE: The ion diameters quoted here are the bare ion diameters. The corresponding hydrated ion diameters vary a lot less. They amount to 0.764 nm (Li+), 0.716 nm (Na+), and 0.658 nm (Cs+). Yet, these values apply only to dilute solutions and would lead to close-packed salt crystals for concentrations well below the 3M in the present experiments.)

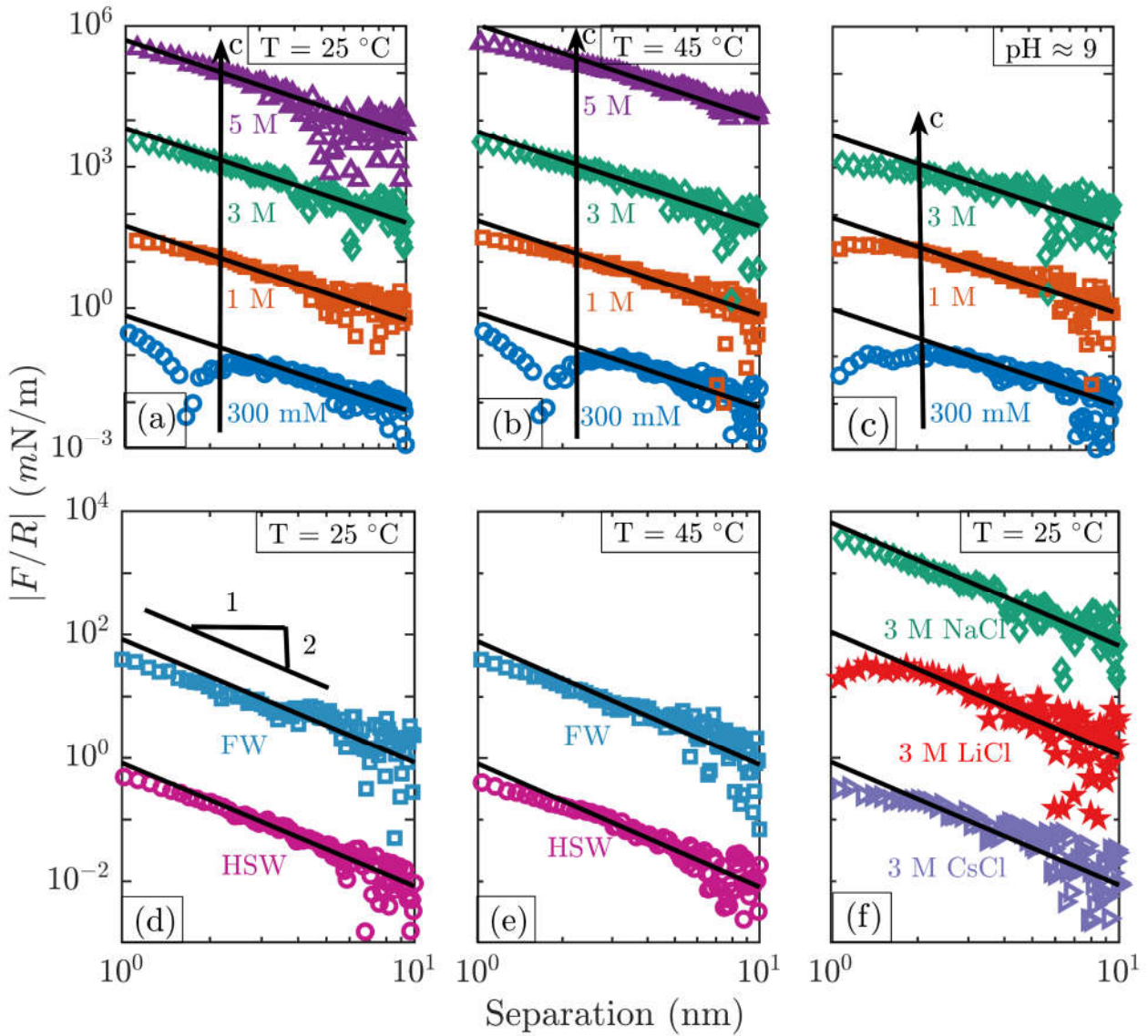

Figure 3: Magnitude of the (overall attractive) normalized force (in high salt concentration regime) vs. tip-sample separation for various conditions: (a) NaCl at 25 °C and unadjusted pH (same data as Fig. 1); (b) NaCl, 45 °C, unadjusted pH; (c) NaCl, 25 °C, pH ≈ 9; (d) High salinity water (HSW) and Formation water (FW), 25 °C, unadjusted pH; (e) HSW and FW, 45 °C, unadjusted pH; (f) various alkali chloride solutions as indicated; conc. 3 M, 25 °C, unadjusted pH with bare-hydrated ionic diameters of 0.188-0.764 nm (Li$^+$), 0.234-0.716 nm (Na$^+$), 0.372-0.658 nm (Cs$^+$). Symbols: experimental data; solid black lines: van der Waals interaction, Eq. (2) with slope -2 in double-logarithmic representation. Force-separation curves are shifted vertically by two orders of magnitude for clarity. Lowest curve unshifted.

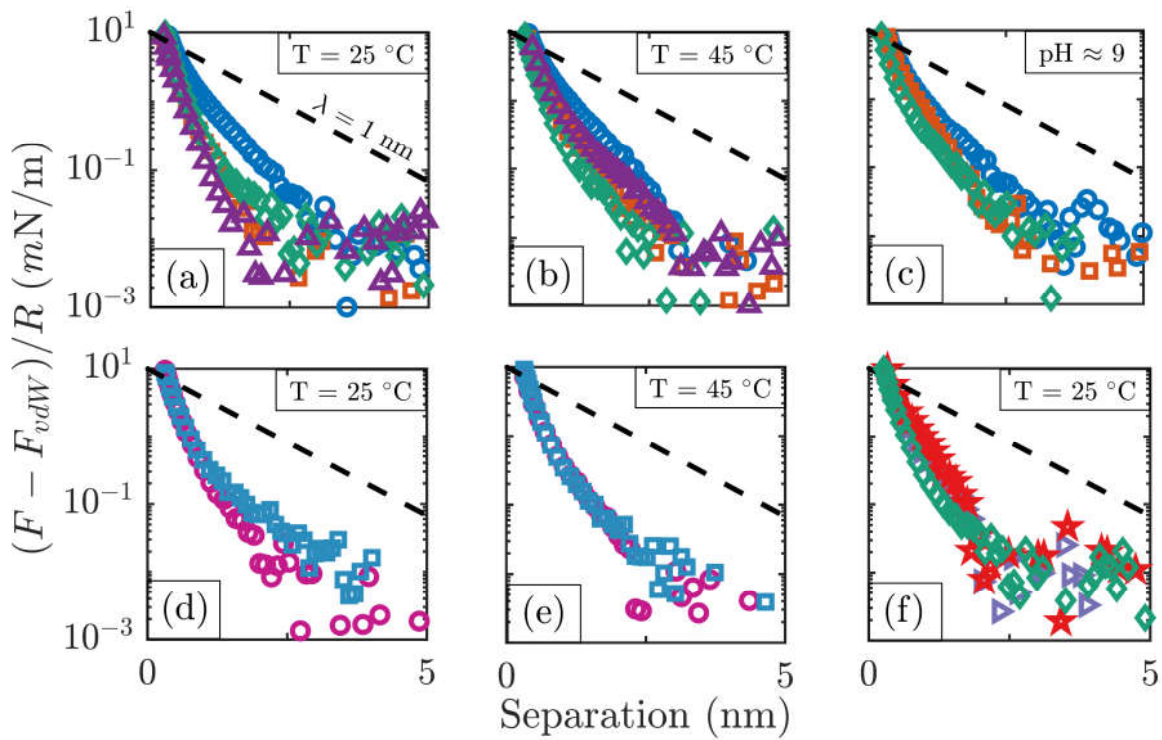

Figure 4: Normalised excess force after subtracting van der Waals interaction vs tip-sample separation (using offset parameter $\delta = 0.1$ nm). Same data and symbols as in Fig. 3. (a) NaCl at 25 °C and unadjusted pH; (b) NaCl, 45 °C , unadjusted pH; (c) NaCl, 25 °C, pH ≈ 9; (d) HSW and FW, 25 °C, unadjusted pH; (e) HSW and FW, 45 °C, unadjusted pH; (f) various alkali chloride solutions as indicated; conc. 3 M, 25 °C, unadjusted pH. Black dashed lines: reference line corresponding to a decay length of 1 nm.

For tip-sample separations below 2 nm, however, we do observe electrolyte-dependent repulsive excess forces, as illustrated by the downward bending of the experimental data away from the van der Waals force in all panels of Figure 3. For the 300 mM NaCl solutions in Figures 3a and 3b, this repulsive force may be explained by a residual continuum electrostatic force, which contributes to a reversal of the total force at a tip-sample separation of $\approx$ 1.5 nm and thereby leads to the sharp local minimum of the curves (see also Figure S5a). For all other electrolytes, these repulsive forces arise from (presumably a combination of one or more of the) non-DLVO forces, as discussed in the literature. To analyze this excess force in more detail, we plot in Figure 4 the deviation of the measured total forces in Figure 3 from the corresponding van der Waals fit on a semi-logarithmic scale. While the exact value of the excess force for $H \to 0$ depends on the (somewhat arbitrary) choice of $\delta$, it is nevertheless clear that the excess force decays very rapidly in all cases. Any possible decay length is shorter than 1nm, as evidenced by comparing to the dashed line in each panel. Within the limited resolution, we cannot identify any clear dependence of this excess force on the salt concentration or brine composition. However, as already stated above, we can exclude the existence of any excess repulsive force with a range beyond 1 nm and a (normalized) strength exceeding 0.1 mN/m.

Combining the decay lengths extracted from the low concentration regime (Figure 2) and the ones estimated based on the excess force between 1 and 2 nm in Figure 4, we find that the experimentally observed decay length of the force (filled colored symbols in Figure 5) decreases with increasing ionic strength following the classical expectations of Debye screening (black solid line) for low salt concentrations up to approximately 0.1 to 1 M and subsequently saturates in a range between 0.1 and 0.5 nm with substantial uncertainties and variations that are limited by experimental resolution rather than any well-controlled physical property of the system. However, what we can exclude for our system is the existence of an excess electrostatic force due to underscreening at high concentrations with a decay length of a few nanometers, as described in the recent SFA experiments. For comparison, we also include experimental data points from ref [21] on concentrated aqueous NaCl solutions (small gray circles) and the corresponding defect model (dashed line) that lead to decay lengths exceeding our experimental observations by more than an order of magnitude upon approach of saturation.

Our numerical DFT calculations confirm the experimentally observed picture. The calculated curves for the shifted surface free energy decay very quickly with increasing tip-sample separation (see Figure S7) and the decay length decreases with increasing ionic strength (crosses in Figure 5). An interesting subtly arises when choosing the (hard sphere) ion size in the DFT calculations. Na+ and Cl- ions have bare ion diameters of $d_+ = 0.234$ nm and $d_- = 0.328$ nm, respectively [48, 49]. However, in aqueous solutions, dissolved ions are dressed by a more or less tightly bound hydration shell and often act as substantially larger entities. For Na+ and Cl-, the hydrated ion diameters are $d_+^h = 0.716$ nm and $d_-^h = 0.664$ nm at infinite dilution and are known to decrease with increasing concentration [50, 51]. By calculating the force curves for both bare and (infinite dilution) hydrated ion diameters we cover the upper and lower limit of any possible intermediate ion diameter for all concentrations. For low concentrations, the resulting decay lengths for both bare (x's in Figure 5) and hydrated (+'s) ion diameters decrease with increasing ionic strength following Debye screening. At higher concentrations, the most important practical consequence is that the calculations for hydrated ions are limited to concentrations below approximately 2 M because of packing constraints: a system with (hard sphere) hydrated ions would thus reach closed packing at a concentration well below the experimental saturation limit and the highest concentration in our experiments. (For the bare ion diameters closed packing is only reached at 25 M.) The choice of the ion radius also has consequences for the location of the Kirkwood point, beyond which the damped oscillatory profiles dominate, and the decay length increases with concentration. For the bare ion diameters, the corresponding critical concentration is ≈1 M, whereas for the hydrated ions, it is close to 0.3 M. As a consequence, a slight increase is seen in the decay length beyond the Kirkwood point for hydrated ions at an ionic strength of 1 M (compared to 0.3 M), in agreement with [27]. Except for this minor effect, the calculations show a consistent decrease of the decay length with increasing ion strength, in agreement with our experimental findings.

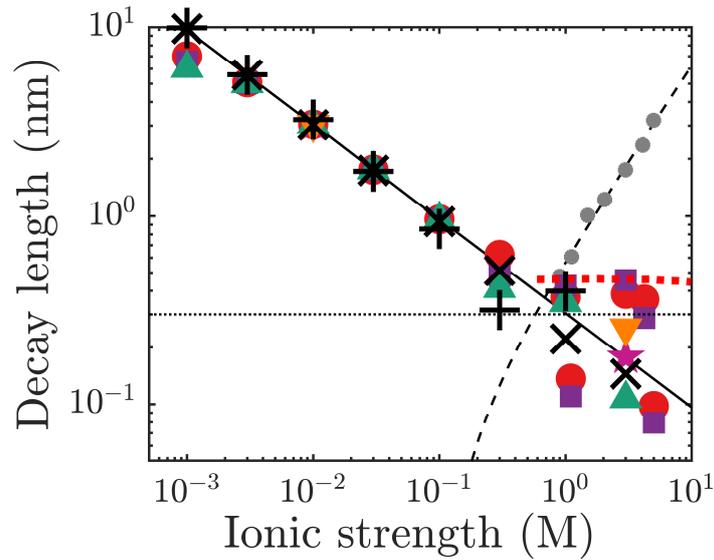

Figure 5: Decay length of repulsive excess force vs. ionic strength. Filled colored symbols: present AFM measurements; ( ○ – NaCl and complex brines of unadjusted pH at 25 °C; □ - NaCl and complex brines of unadjusted pH at 45 °C; Δ - NaCl at 25 °C and pH ≈ 9; ▽ – CsCl of unadjusted pH at 25 °C; ☆ - LiCl of unadjusted pH at 25 °C.) Crosses: DFT calculations for bare ion diameters (x) and hydrated ion diameters (+). The symbol in the braces present maximum packing for the DFT calculation. Dotted red line: model from ref. [26] . Solid blackline: Debye screening length. Thin dotted black line: diameter of $H_2O$ molecules. Gray circles and gray dashed line: results from ref [21].

**Discussion**

The key result of our experiments and calculations is thus that we do not find any evidence for anomalous underscreening between silica surfaces in aqueous salt solutions across a wide range of salt concentrations and ionic radii. In the experiments, we explored strongly and weakly hydrated cations as well as mixtures of ions. We also varied temperature and pH of the solution in our experiments such that we covered a range of bulk dielectric constants from ~74(300 mM) to ~43(5 M) at 25 °C and from ~67(300 mM) to 32(5 M) at 45 °C[52], which affects the Bjerrum length. The consistently sub-nanometric decay lengths in AFM experiments over such a wide range of conditions leads us to the conclusion that anomalous underscreening should most likely not be considered a universal intrinsic property of concentrated aqueous electrolytes. Instead, our experiments with common very smooth silica surfaces suggest that screening of electrostatic forces simply improves with increasing salt concentration until short-range forces take over at distances below 1 nm, in agreement with the DFT calculations of the primitive model.

If underscreening is indeed not a universal phenomenon of concentrated electrolyte, the question arises whether the deviation between the present AFM experiments and the SFA measurements are either caused by specific properties of the surfaces (mica vs. silica) or by the specific configuration and measurement protocols of the two types of instruments. The resolution of both methods in terms of normalized forces ($F/R$) is comparable. In particular, the magnitude of the long-range monotonically decaying force in ref. [20] of 0.1 - 1 N/m clearly falls within the resolution of our colloidal probe AFM measurements. Similarly, the Péclet number $Pe = u\sqrt{RH}/D = O(10^{-6} \ldots 10^{-5})$ ($u$: approach speed; $D$: diffusion coefficient), which describes the ratio between advective and diffusive transport, suggests that both types of measurements should report mechanically equilibrated forces. (Unfortunately, though, the agreement between approach and retract curves – see inset of Figure 1 – which is a prerequisite for any meaningful comparison to calculated equilibrium forces, is hardly ever demonstrated in SFA studies.)

The most important remaining difference thus arises from the different substrate materials. The silica surfaces in the present study display a small but finite roughness, which is comparable to the size of water molecules (see Figure S2). This is presumably the reason for the absence of oscillatory forces in our measurement. Note, however that the residual surfaces roughness is nevertheless small compared to the expected length scale of the asymptotically decaying long range force reported in the literature [19-21]. Silica surfaces also carry less intrinsic surface charge than mica. Yet, the very high intrinsic surface charge density on mica is usually compensated within the Stern layer at mica-electrolyte interfaces, resulting in diffuse-layer potentials, zeta potentials and corresponding diffuse-layer charge densities – that are comparable to silica, certainly at slightly elevated pH, as in Figure 3c and 4c[8, 53, 54]. On the other hand, the combination of very high intrinsic charge density and crystalline structure does induce a number of specific effects at mica-electrolyte interfaces. For instance, unlike silica and most other materials, mica-electrolyte interfaces can display overcharging even in solutions with monovalent cations. This has been shown explicitly in x-ray surface diffraction measurements for RbCl [55] and CsCl [56] solutions and – albeit implicitly – in AFM measurements [57]. The x-ray experiments also show strong evidence of a layer of $Cl^-$ anions on top of the first layer of $Rb^+$ ions suggesting a transition towards an ionic crystal at concentrations beyond a few hundred mM. Such ionic nano-crystals have indeed been seen in MD simulations of RbI solutions at concentrations that were high, but still within the range of stable bulk solutions [55]. In all cases, the specific ionic structures at the interfaces are stabilized by the characteristic geometry and (charged) sites on the mica surface. Possibly, the same peculiarities of mica are also responsible for the long-range density oscillations in AFM measurements in slowly evaporating KCl solutions upon approaching precipitation [58]. In exploratory force measurements with 3 M NaCl solutions confined between a mica surface and the same type of silica probe used above we found a very similar behavior as in the silica-silica system (see Figure S8) i.e., no anomalous screening length. Yet, the situation may well be different in case of two confining crystalline mica surfaces, as in the SFA.

**Conclusion**

Our AFM measurements and our classical DFT calculations of tip-sample interaction forces in aqueous electrolytes of variable composition (concentration, cation species, pH, temperature) and complexity follow the expectations based on classical DLVO theory for tip-sample separations of 2 nm and more. For salt concentrations up to 100 mM, the electrostatic interactions decay exponentially with a decay length that decreases with increasing ionic strength, in accordance with the definition of the Debye screening length. For ionic strengths of 300 mM and beyond, electrostatic forces are efficiently screened and the total tip-sample interaction force at long distances is governed by attractive van der Waals forces ($\propto H^{-2}$). For tip-sample separation below 2 nm, an additional excess repulsive force is observed that decreases rapidly with a decay length of less than 1 nm for all fluid compositions investigated. In none of the fluid compositions that we investigated did we observe an increase in range of the excess force with increasing salt concentration as would be indicative of the recently reported anomalous underscreening. Accordingly, neither the measured decay lengths of the excess force nor the calculated ones follow the proposed scaling of underscreening with ion size and Bjerrum length. Together, our experiments and calculations consistently show that the phenomenon of anomalous underscreening is not as universal as perhaps presumed earlier. Based on the wide range of fluid compositions investigated we feel confident in claiming that anomalous underscreening does not occur for silica surfaces and confinement geometries as encountered in an AFM. Consequently, the phenomenon does not seem to be an intrinsic equilibrium property of highly concentrated electrolytes but rather depends on interfacial properties and/or geometric confinements.

**Acknowledgement:** SK, IS and FM would like to thank University of Twente and the Saudi Arabian Oil Company (Saudi Aramco) for the funding under the contract no. 6600041100. PC and RR would like to thank the D-ITP consortium, a program of the Netherlands Organisation for Scientific Research (NWO) that is funded by the Dutch Ministry of Education, Culture and Science (OCW)

# References

1. Verwey, E.J.W., J.T.G. Overbeek, and K. Van Nes, *Theory of the stability of lyophobic colloids: the interaction of sol particles having an electric double layer*. 1948: Elsevier Publishing Company.
2. Ninham, B., *Selected works of BV Deryaguin.* Progress in Surface Science, 1994. **47**(4): p. 5-8.
3. Israelachvili, J.N., *Intermolecular and surface forces*. 2015: Academic press.
4. Ninham, B.W., *On progress in forces since the DLVO theory.* Advances in colloid and interface science, 1999. **83**(1-3): p. 1-17.
5. Ben-Yaakov, D., et al., *Beyond standard Poisson–Boltzmann theory: ion-specific interactions in aqueous solutions.* Journal of Physics: Condensed Matter, 2009. **21**(42): p. 424106.
6. Ben-Yaakov, D., et al., *Ion-specific hydration effects: Extending the Poisson-Boltzmann theory.* Current Opinion in Colloid & Interface Science, 2011. **16**(6): p. 542-550.
7. Bourg, I.C., et al., *Stern layer structure and energetics at mica–water interfaces.* The Journal of Physical Chemistry C, 2017. **121**(17): p. 9402-9412.
8. Van Lin, S.R., et al., *Ion-specific and ph-dependent hydration of mica–electrolyte interfaces.* Langmuir, 2019. **35**(17): p. 5737-5745.
9. Chapel, J.-P., *Electrolyte species dependent hydration forces between silica surfaces.* Langmuir, 1994. **10**(11): p. 4237-4243.
10. Voïtchovsky, K., et al., *Direct mapping of the solid–liquid adhesion energy with subnanometre resolution.* Nature Nanotechnology, 2010. **5**(6): p. 401-405.
11. Fukuma, T., et al., *Atomic-scale distribution of water molecules at the mica-water interface visualized by three-dimensional scanning force microscopy.* Physical review letters, 2010. **104**(1): p. 016101.
12. Söngen, H., et al., *Resolving point defects in the hydration structure of calcite (10.4) with three-dimensional atomic force microscopy.* Physical review letters, 2018. **120**(11): p. 116101.
13. Simon, P., Y. Gogotsi, and B. Dunn, *Where do batteries end and supercapacitors begin?* Science, 2014. **343**(6176): p. 1210-1211.
14. Diao, Y. and R.M. Espinosa-Marzal, *The role of water in fault lubrication.* Nature communications, 2018. **9**(1): p. 1-10.
15. Eisenberg, B., *Ionic interactions are everywhere.* Physiology, 2013. **28**(1): p. 28-38.
16. Armand, M. and J.-M. Tarascon, *Building better batteries.* nature, 2008. **451**(7179): p. 652-657.
17. Lu, Q., et al., *A selective and efficient electrocatalyst for carbon dioxide reduction.* Nature communications, 2014. **5**(1): p. 1-6.
18. Lake, L.W., *Enhanced oil recovery.* 1989.

19. Baimpos, T., et al., *Effect of interfacial ion structuring on range and magnitude of electric double layer, hydration, and adhesive interactions between mica surfaces in 0.05–3 M Li+ and Cs+ electrolyte solutions.* Langmuir, 2014. **30**(15): p. 4322-4332.
20. Smith, A.M., A.A. Lee, and S. Perkin, *The electrostatic screening length in concentrated electrolytes increases with concentration.* The journal of physical chemistry letters, 2016. **7**(12): p. 2157-2163.
21. Perez-Martinez, C.S., A.M. Smith, and S. Perkin, *Scaling analysis of the screening length in concentrated electrolytes.* Physical review letters, 2017. **119**(2): p. 026002.
22. Perez-Martinez, C.S., A.M. Smith, and S. Perkin, *Underscreening in concentrated electrolytes.* Faraday discussions, 2017. **199**: p. 239-259.
23. Smith, A.M., et al., *Unexpectedly Large Decay Lengths of Double-Layer Forces in Solutions of Symmetric, Multivalent Electrolytes.* The Journal of Physical Chemistry B, 2019. **123**(7): p. 1733-1740.
24. Gebbie, M.A., et al., *Ionic liquids behave as dilute electrolyte solutions.* Proceedings of the National Academy of Sciences, 2013. **110**(24): p. 9674-9679.
25. Gebbie, M.A., et al., *Long-range electrostatic screening in ionic liquids.* Proceedings of the National Academy of Sciences, 2015. **112**(24): p. 7432-7437.
26. Rotenberg, B., O. Bernard, and J.-P. Hansen, *Underscreening in ionic liquids: a first principles analysis.* Journal of Physics: Condensed Matter, 2018. **30**(5): p. 054005.
27. Cats, P., et al., *Primitive model electrolytes in the near and far field: Decay lengths from DFT and simulations.* The Journal of Chemical Physics, 2021. **154**(12): p. 124504.
28. Burak, Y. and D. Andelman, *Hydration interactions: Aqueous solvent effects in electric double layers.* Physical Review E, 2000. **62**(4): p. 5296.
29. Hutter, J.L. and J. Bechhoefer, *Calibration of atomic‐force microscope tips.* Review of Scientific Instruments, 1993. **64**(7): p. 1868-1873.
30. Carnie, S.L. and D.Y. Chan, *Interaction free energy between plates with charge regulation: a linearized model.* Journal of colloid and interface science, 1993. **161**(1): p. 260-264.
31. Zhao, C., et al., *Extracting local surface charges and charge regulation behavior from atomic force microscopy measurements at heterogeneous solid-electrolyte interfaces.* Nanoscale, 2015. **7**(39): p. 16298-16311.
32. Behrens, S.H. and M. Borkovec, *Electric double layer interaction of ionizable surfaces: Charge regulation for arbitrary potentials.* The Journal of chemical physics, 1999. **111**(1): p. 382-385.
33. Valmacco, V., et al., *Forces between silica particles in the presence of multivalent cations.* Journal of colloid and interface science, 2016. **472**: p. 108-115.
34. Parsegian, V.A., *Van der Waals forces: a handbook for biologists, chemists, engineers, and physicists*. 2005: Cambridge University Press.
35. Ackler, H.D., R.H. French, and Y.-M. Chiang, *Comparisons of Hamaker constants for ceramic systems with intervening vacuum or water: From force laws and physical properties.* Journal of colloid and interface science, 1996. **179**(2): p. 460-469.
36. Bergström, L., *Hamaker constants of inorganic materials.* Advances in colloid and interface science, 1997. **70**: p. 125-169.

37. Evans, R., *The nature of the liquid-vapour interface and other topics in the statistical mechanics of non-uniform, classical fluids.* Advances in physics, 1979. **28**(2): p. 143-200.
38. Hansen-Goos, H. and R. Roth, *Density functional theory for hard-sphere mixtures: the White Bear version mark II.* Journal of Physics: Condensed Matter, 2006. **18**(37): p. 8413.
39. Roth, R., *Fundamental measure theory for hard-sphere mixtures: a review.* Journal of Physics: Condensed Matter, 2010. **22**(6): p. 063102.
40. Mier‐y‐Teran, L., et al., *A nonlocal free‐energy density‐functional approximation for the electrical double layer.* The Journal of Chemical Physics, 1990. **92**(8): p. 5087-5098.
41. Cats, P., et al., *Capacitance and Structure of Electric Double Layers: Comparing Brownian Dynamics and Classical Density Functional Theory.* arXiv preprint arXiv:2103.00897, 2021.
42. Bolt, G., *Determination of the charge density of silica sols.* The journal of physical chemistry, 1957. **61**(9): p. 1166-1169.
43. Revil, A., et al., *Streaming potential in porous media: 2. Theory and application to geothermal systems.* Journal of Geophysical Research: Solid Earth, 1999. **104**(B9): p. 20033-20048.
44. Klaassen, A., et al., *Impact of surface defects on the surface charge of gibbsite nanoparticles.* Nanoscale, 2017. **9**(14): p. 4721-4729.
45. Ducker, W.A., T.J. Senden, and R.M. Pashley, *Measurement of forces in liquids using a force microscope.* Langmuir, 1992. **8**(7): p. 1831-1836.
46. Grabbe, A. and R.G. Horn, *Double-layer and hydration forces measured between silica sheets subjected to various surface treatments.* Journal of colloid and interface science, 1993. **157**(2): p. 375-383.
47. Liu, F., *A study of interaction forces at the solid-liquid interface using atomic force microscopy.* 2016.
48. Lide, D.R., *CRC handbook of chemistry and physics*. Vol. 85. 2004: CRC press.
49. Shannon, R.D., *Revised effective ionic radii and systematic studies of interatomic distances in halides and chalcogenides.* Acta crystallographica section A: crystal physics, diffraction, theoretical and general crystallography, 1976. **32**(5): p. 751-767.
50. Marcus, Y., *Concentration dependence of ionic hydration numbers.* The Journal of Physical Chemistry B, 2014. **118**(35): p. 10471-10476.
51. Espinosa-Marzal, R.M., et al., *Hydrated-ion ordering in electrical double layers.* Physical Chemistry Chemical Physics, 2012. **14**(17): p. 6085-6093.
52. Buchner, R., G.T. Hefter, and P.M. May, *Dielectric relaxation of aqueous NaCl solutions.* The Journal of Physical Chemistry A, 1999. **103**(1): p. 1-9.
53. Dove, P.M. and C.M. Craven, *Surface charge density on silica in alkali and alkaline earth chloride electrolyte solutions.* Geochimica et Cosmochimica Acta, 2005. **69**(21): p. 4963-4970.
54. Sposito, G., *The surface chemistry of natural particles*. 2004: Oxford University Press on Demand.
55. Lee, S.S., et al., *Ion correlations drive charge overscreening and heterogeneous nucleation at solid–aqueous electrolyte interfaces.* Proceedings of the National Academy of Sciences, 2021. **118**(32).

56. Brugman, S.J., et al., *Concentration-dependent adsorption of CsI at the muscovite–electrolyte interface.* Langmuir, 2018. **34**(13): p. 3821-3826.
57. Ricci, M., P. Spijker, and K. Voïtchovsky, *Water-induced correlation between single ions imaged at the solid–liquid interface.* Nature communications, 2014. **5**(1): p. 1-8.
58. Martin-Jimenez, D., et al., *Atomically resolved three-dimensional structures of electrolyte aqueous solutions near a solid surface.* Nature communications, 2016. **7**(1): p. 1-7.